# Note on the Lattice Fermion Chiral Symmetry Group

Jeffrey E. Mandula[*]

University of Washington, Department of Physics, Seattle, WA 98105, USA

The group structure of the variant chiral symmetry discovered by Lüscher in the Ginsparg-Wilson description of lattice chiral fermions is analyzed. It is shown that the group contains an infinite number of linearly independent symmetry generators, and the Lie algebra is given explicitly. CP is an automorphism of the chiral group, and the CP transformation properties of the symmetry generators is found. Features of the currents associated with these symmetries are discussed, including the fact that some different, non-commuting symmetry generators lead to the same Noether current. These strange features occur in all implementations of lattice fermions based on the Ginsparg-Wilson relation, including overlap, domain-wall, and perfect-action chiral fermions. The conclusions are illustrated in a solvable example – free overlap fermions.

---

[*] E-mail address: mandula@post.harvard.edu

The close of the 1990's saw a very largely successful effort to represent chiral fermions on the lattice via several constructions [1,2,3], all of which, it was realized after the fact, crucially incorporated the Ginsparg-Wilson relation [4]. Lüscher observed that there was a transformation on the fermion variables that, as a consequence of the Ginsparg-Wilson relation, left the fermion action invariant [5]. This variant of chiral symmetry has several unusual features: fermion and anti-fermion variables are treated asymmetrically, the effect of the transformations of the fermion variables explicitly depends on the lattice gauge variables, and the symmetry can be expressed in multiple mathematically inequivalent (though presumably physically equivalent) ways.

The purpose of this note is to point out several problems that are consequences of the last feature. They relate to the structure of the lattice chiral group and the connection between the group generators and their currents. The three elements that we will examine are: 1) the structure of the lattice chiral group, which is quite elaborate and quite different than in the continuum; 2) the action of CP transformations on the generators of this group, which may clarify the difficulties experienced in representing chiral interactions using Ginsparg-Wilson fermions; and 3) a mismatch between Noether currents of the symmetries and the symmetries themselves, which highlights non-canonical aspects of the Euclidean path integral.

The problems discussed in this note are common to all implementations of lattice fermions based on the Ginsparg-Wilson relation — overlap, domain-wall, and perfect-action chiral fermions.

*Group Structure:* The following discussion uses only the Ginsparg-Wilson relation [4], the most familiar form of which (suppressing the lattice spacing) is

$$\gamma_5 D + D \gamma_5 = D \gamma_5 D \tag{1}$$

where $D$ is the lattice Dirac kernel, a $\gamma_5$-hermitian ($\gamma_5 D \gamma_5 = D^\dagger$) matrix labeled by color, spin, and lattice site indices. If we write the kernel as $D = 1 - V$, then $V$ is unitary (as well as $\gamma_5$-hermitian). In fact, these properties of $V$,

$$V^{-1} = V^\dagger = \gamma_5 V \gamma_5 \tag{2}$$

are completely equivalent to Ginsparg-Wilson equation (1).

The chiral transformations under which the fermion action $S_F = \bar\psi D \psi$ is invariant are generated by [5]

$$\begin{aligned}\delta \psi &= \gamma_5 (1 - D) \psi = \gamma_5 V \psi \\ \delta \bar\psi &= \bar\psi \gamma_5\end{aligned} \tag{3}$$

The asymmetric treatment of lattice fermions and anti-fermions is allowed because fermions and anti-fermions enter the Euclidean path integral as independent, not conjugate, variables. One may modify the transformation of the anti-fermion variables instead of the fermion variables,

$$\begin{aligned}\delta' \psi &= \gamma_5 \psi \\ \delta' \bar\psi &= \bar\psi (1 - D) \gamma_5 = \bar\psi V \gamma_5\end{aligned} \tag{4}$$

or use any linear combination of these transformation rules. Whatever one's choice, the Ginsparg-Wilson equation insures the invariance of the action $S_F$.

These transformations are universally regarded as physically equivalent, but they are certainly different. Furthermore, each generates a symmetry of $S_F$. This means that the full chiral symmetry group of this theory is not generated by a choice of one or the other of these transformations (nor a linear combination thereof), but by both. This "enlarged" chiral group is not the end of the story either, because the two transformations do not even commute. Explicitly,

$$[\delta', \delta]\psi = \delta'\delta\psi - \delta\delta'\psi = (V^{-1} - V)\psi$$
$$[\delta', \delta]\bar{\psi} = \delta'\delta\bar{\psi} - \delta\delta'\bar{\psi} = \bar{\psi}(V - V^{-1})$$
(5)

Their commutator, a vector rather than an axial transformation, is yet another symmetry of the fermion action. Upon further commutation, each of the chiral transformations (3) and (4) generates new symmetry transformations, further enlarging the chiral group.

The full group is easily found by writing the Ginsparg-Wilson equation as

$$\gamma_5 D + D\gamma_5 V = 0 \qquad (6)$$

a form that displays the symmetry under Eq. (3) most clearly. Right multiplication by $V^{n-1}$ and the use of the unitarity and $\gamma_5$-hermiticity of $V$ gives

$$V^{1-n}\gamma_5 D + D\gamma_5 V^n = 0 \qquad (7)$$

while right multiplication by $\gamma_5 V^{n+1}$, and noting that $\gamma_5 D\gamma_5 = -V^{-1}D$, gives

$$-V^n D + DV^n = 0 \qquad (8)$$

From these we can read off the remaining axial and vector transformations of the fermion variables that generate symmetries of the Ginsparg-Wilson action. They are

$$\delta_A^{(n)} \psi = \gamma_5 V^n \psi$$
$$\delta_A^{(n)} \bar{\psi} = \bar{\psi} V^{1-n} \gamma_5$$

$$\delta_V^{(n)} \psi = -i V^n \psi$$
$$\delta_V^{(n)} \bar{\psi} = i \bar{\psi} V^n$$
(9)

We recognize the two axial transformations with which we began this discussion as

$$\delta = \delta_A^{(1)}$$
$$\delta' = \delta_A^{(0)}$$
(10)

The commutators of the generalized axial and vector generators are

$$[\delta_A^{(n)}, \delta_A^{(m)}] = i\left(\delta_V^{(n-m)} - \delta_V^{(m-n)}\right)$$
$$[\delta_V^{(n)}, \delta_A^{(m)}] = i\left(\delta_A^{(m-n)} - \delta_A^{(m+n)}\right)$$
$$[\delta_V^{(n)}, \delta_V^{(m)}] = 0$$
(11)

This shows that the algebra closes and so we have identified the full chiral group. It has infinite rank, and the vector transformations $\delta_V^{(n)}$ are mutually commuting generators.

The axial generators carrying definite values of these vector generators are linear combinations of the $\delta_A^{(m)}$. Since Eq. (11) (specifically the second line) has the form of a constant coefficient linear difference equation for $\delta_A^{(m)}$, the eigenvectors of $\delta_V^{(n)}$ under commutation are Fourier series in $\delta_A^{(m)}$:

$$\delta_A(\phi) = \sum_m e^{-im\phi} \delta_A^{(m)}$$
(12)

Their eigenvalues under commutation are shown by

$$[\delta_V^{(n)}, \delta_A(\phi)] = 2\sin n\phi \, \delta_A(\phi)$$
(13)

These linear combinations of generators, $\delta_A(\phi)$, raise and lower the eigenvalues of each of the diagonal generators $\delta_V^{(n)}$.

We have suppressed flavor symmetry in this discussion, but it may be reinstated by modifying the transformation rules Eq. (9) to include the appropriate flavor $\lambda$-matrices. The only interesting change this brings is that with flavor, the vector transformations do not commute among themselves and each of them appears individually on the right-hand side of the symmetry algebra, as opposed to just the combinations $\delta_V^{(n)} - \delta_V^{(-n)}$.

Infinite parameter symmetry groups are often a sign of some disease in a theory, and so a few remarks about this are in order. First of all, the elaborate structure of the lattice chiral symmetry group that we have described is built into the Ginsparg-Wilson equation, and applies to all implementations of the relation. The enlargement of the symmetry group, in fact the existence of the group itself, depends on $\psi$ and $\bar\psi$ being independent variables in the Euclidean space path integral.

*CP Symmetry:* The only notable failure of the Ginsparg-Wilson implementation of chiral symmetry has been the inability, at least to date, to use it to construct lattice fermions with chiral interactions [6]. This failure is related to the fact that CP transforms the two forms of the asymmetric chiral transformations $\delta$ and $\delta'$ into one another. However, we will see that CP is an automorphism of the enlarged chiral group.

Under parity, the fermion variables and $V = 1 - D$ transform as

$$\mathcal{P} \, \psi \to \gamma_4 P \psi$$
$$\mathcal{P} \, \bar{\psi} \to \bar{\psi} P \gamma_4 \qquad (14)$$
$$\mathcal{P} \, V \to \gamma_4 P V P \gamma_4$$

where the matrix $P = P^{-1}$ acts only on the site labels and reflects the $x_{1,2,3}$ indices.

Charge conjugation is represented on the link variables by complex conjugation, so the fermion variables and $V$ transform as

$$\mathcal{C} \, \psi \to C \, \bar{\psi}^{\mathsf{T}}$$
$$\mathcal{C} \, \bar{\psi} \to -\psi^{\mathsf{T}} C^{-1} \qquad (15)$$
$$\mathcal{C} \, V \to \Gamma V^* \Gamma^{-1}$$

The matrices $C$ and $\Gamma$ depend on the representation of the Dirac matrices. $\Gamma$ flips the sign of the imaginary $\gamma$-matrices,

$$\Gamma \gamma_\mu^* \Gamma^{-1} = \gamma_\mu \qquad (16)$$

and invariance of the action requires $C = \gamma_5 \Gamma$.

The CP transformation properties of the generators of the lattice chiral group follow from their action on the CP transformed variables. Straightforward algebra gives

$$(\mathcal{CP})^{-1} \, \delta_A^{(n)} \, (\mathcal{CP}) \, \psi = -\gamma_5 V^{1-n} \psi$$
$$(\mathcal{CP})^{-1} \, \delta_A^{(n)} \, (\mathcal{CP}) \, \bar{\psi} = -\bar{\psi} V^n \gamma_5$$
$$\qquad (17)$$
$$(\mathcal{CP})^{-1} \, \delta_V^{(n)} \, (\mathcal{CP}) \, \psi = i V^n \psi$$
$$(\mathcal{CP})^{-1} \, \delta_V^{(n)} \, (\mathcal{CP}) \, \bar{\psi} = -i\bar{\psi} V^n$$

Comparing the above with the definitions of the generators (9) shows that

$$(\mathcal{CP})^{-1} \, \delta_A^{(n)} \, (\mathcal{CP}) = -\delta_A^{(1-n)}$$
$$(\mathcal{CP})^{-1} \, \delta_V^{(n)} \, (\mathcal{CP}) = -\delta_V^{(n)} \qquad (18)$$

Thus CP is seen to be an automorphism of the extended chiral symmetry group, but one that mixes the axial generators. This is the symmetry group view of the difficulty in reconciling CP invariance with chiral symmetry as realized through the Ginsparg-Wilson equation.

*Currents:* The conserved currents associated with each of the $\delta_A^{(n)}$ and $\delta_V^{(n)}$ symmetries are constructed following the Noether procedure. We take $\varepsilon(x)$ to be an infinitesimal function of lattice site and denote the matrix that projects on site $x$ by $(I(x))_{y,z} = \delta_{y,x}\delta_{x,z}$. Let $E$ to be the diagonal matrix

$$E = \sum_x \varepsilon(x) I(x) \tag{19}$$

The variations of the fermion variables under space-time varying versions of each of the axial and vector symmetries of the fermion action are then generated by

$$\begin{aligned}
\Delta_A^{(n)} \psi &= E\, \gamma_5 V^n \psi \\
\Delta_A^{(n)} \bar\psi &= \bar\psi\, V^{1-n} \gamma_5 E \\
\\
\Delta_V^{(n)} \psi &= -i E\, V^n \psi \\
\Delta_V^{(n)} \bar\psi &= i \bar\psi\, V^n E
\end{aligned} \tag{20}$$

and so their effects on the fermionic action $S_F = \bar\psi(1-V)\psi$ are

$$\begin{aligned}
\Delta_A^{(n)} S_F &= \bar\psi\{(1-V) E\, \gamma_5 V^n + V^{1-n} \gamma_5 E(1-V)\}\psi \\
\Delta_V^{(n)} S_F &= -i\bar\psi\{(1-V) E\, V^n \;-\; V^n E(1-V)\}\psi
\end{aligned} \tag{21}$$

The conserved currents are the coefficients of $\partial_\mu^{(+)}\varepsilon(x)$ in each of these expressions,

$$\begin{aligned}\Delta_A^{(n)} S_F &= \sum_{x,\mu} \left(\partial_\mu^{(+)} \varepsilon(x)\right) J_\mu^{5(n)}(x) \\ \Delta_V^{(n)} S_F &= \sum_{x,\mu} \left(\partial_\mu^{(+)} \varepsilon(x)\right) J_\mu^{(n)}(x)\end{aligned} \qquad (22)$$

where $\partial_\mu^{(+)}$ is the forward, nearest neighbor ordinary (as opposed to covariant) difference.

From the above expressions for the lattice-site dependent variations of the action, we note a very curious property of the currents associated with the symmetries with which we began this discussion, the axial transformations generated by $\delta_A^{(1)}$ and by $\delta_A^{(0)}$, and also of the vector transformations generated by $\delta_V^{(1)}$ and by $\delta_V^{(0)}$. Even though the two pairs of transformations are different, their currents are the same! As one can see from Eq. (21),

$$\begin{aligned}\Delta_A^{(0)} S_F &= \Delta_A^{(1)} S_F \\ \Delta_V^{(0)} S_F &= \Delta_V^{(1)} S_F\end{aligned} \qquad (23)$$

and so the associated currents must be equal.

$$\begin{aligned}J_\mu^{5(0)}(x) &= J_\mu^{5(1)}(x) \\ J_\mu^{(0)}(x) &= J_\mu^{(1)}(x)\end{aligned} \qquad (24)$$

In a canonical treatment of any field theory, these results would be impossible. In canonically quantized field theory, the generators of symmetry transformations are the space integrals of the time components of their conserved currents. Since the same currents are associated with different transformations, the process of recovering a valid Minkowski space theory from Euclidean space Ginsparg-Wilson fermions must somehow cause these transformations to coalesce.

There is a second aspect of this observation, which is just as remarkable. Each of the symmetry transformations generated by $\delta_A^{(n)}$ and $\delta_V^{(n)}$ is a global transformation, one whose magnitude is the same at all sites. The associated site-dependent transformations generated by $\Delta_A^{(n)}$ and $\Delta_V^{(n)}$ are local transformations but not symmetries, *except for two linear combinations*:

$$\begin{aligned}\left(\Delta_A^{(0)} - \Delta_A^{(1)}\right)S_F &= 0 \\ \left(\Delta_V^{(0)} - \Delta_V^{(1)}\right)S_F &= 0\end{aligned} \tag{25}$$

These two linear combinations are the generators of local gauge symmetries. However, they are quite unlike usual gauge symmetries, in that these transformation act only on the fermion variables, and leave the lattice gauge variables unchanged.

There are, of course, many algebraic relations between the constant and space-time varying symmetry generators. From their actions on the fermionic and anti-fermionic variables, one finds

$$\begin{aligned}\delta_A^{(n)}\Delta_A^{(m)} - \delta_A^{(m)}\Delta_A^{(n)} &= i\left(\Delta_V^{(n-m)} - \Delta_V^{(m-n)}\right) \\ \delta_A^{(n)}\left(\Delta_V^{(m)} - \Delta_V^{(-m)}\right) &= i\left(\Delta_A^{(n+m)} - \Delta_A^{(n-m)}\right) \\ \left(\delta_V^{(m)} - \delta_V^{(-m)}\right)\Delta_A^{(n)} &= -i\left(\Delta_A^{(n+m)} - \Delta_A^{(n-m)}\right)\end{aligned} \tag{26}$$

Since the currents are mechanically extracted from $\Delta_{A,V}^{(n)} S_F$, these are likewise symmetry relations between the Noether currents. The space-time dependent transformations reduce to the symmetry generators when $\varepsilon(x)$ is constant. Therefore making the replacement $\Delta_{A,V}^{(n)} \to \delta_{A,V}^{(n)}$ in Eq. (26) gives a set of relations between the global symmetry generators. These are, of course, consistent with the commutation relations (11).

We can explicitly display the currents using the method of Kikukawa and Yamada [7], which employs the overlap formulation of chiral fermions, starting from the Wilson fermion action. We first put the site-dependent variations of the action (21) into a form in which the matrix $E$ appears only in commutators.

$$\Delta_A^{(n)} S_F = \bar{\psi} \gamma_5 \left\{ [E, V^{n-1}](V-1) - V^{-1}[E,V]V^{n-1} \right\} \psi$$
$$\Delta_V^{(n)} S_F = i \bar{\psi} \left\{ V[E, V^{n-1}](V-1) - [E,V]V^{n-1} \right\} \psi \tag{27}$$

We then expand the commutators using

$$[E, V^{n-1}] = \sum_{m=0}^{n-2} V^m [E,V] V^{n-m-2} \qquad (n \geq 2)$$
$$= -\sum_{m=1}^{1-n} V^{-m} [E,V] V^{n+m-2} \qquad (n \leq 0) \tag{28}$$

which converts the expressions for the variations of the action to sums of terms in which $E$ enters only through the single commutator $[E,V]$.

For the overlap kernel based on Wilson fermions, Kikukawa and Yamada [7] have evaluated $[E,V]$. In a matrix notation, their result is

$$[E,V] = \sum_{x,\mu} \left( \partial_\mu^{(+)} \varepsilon(x) \right) K_\mu(x)$$
$$K_\mu(x) = \frac{1}{\pi} \int_{-\infty}^{+\infty} \frac{dt}{t^2 + D_W^\dagger D_W} \left\{ T_\mu(x) t^2 + D_W T_\mu^\dagger(x) D_W \right\} \frac{1}{t^2 + D_W^\dagger D_W} \tag{29}$$
$$T_\mu(x) = \frac{1}{2} \left[ (\gamma_\mu - 1) I(x) U_\mu + (\gamma_\mu + 1) U_\mu^\dagger I(x) \right]$$

where $D_W$ is the Wilson fermion kernel and $(U_\mu)_{x,y} = \delta_{x+\hat{\mu},y} U_\mu(x)$ is the matrix of lattice link variables between sites.

The currents are then obtained from the expanded forms of Eq. (27) simply by replacing each occurrence of $[E,V]$ by $K_\mu(x)$, giving the rather ungainly results

$$\begin{aligned}
J_\mu^{5(n)}(x) &= \bar{\psi}\gamma_5 \left\{ \sum_{m=0}^{n-2} V^m K_\mu(x) V^{n-m-2}(V-1) - V^{-1} K_\mu(x) V^{n-1} \right\} \psi & (n \geq 2) \\
&= \bar{\psi}\gamma_5 \left\{ -\sum_{m=1}^{1-n} V^{-m} K_\mu(x) V^{n+m-2}(V-1) - V^{-1} K_\mu(x) V^{n-1} \right\} \psi & (n \leq 0) \\
J_\mu^{(n)}(x) &= i\bar{\psi} \left\{ V\sum_{m=0}^{n-2} V^m K_\mu(x) V^{n-m-2}(V-1) - K_\mu(x) V^{n-1} \right\} \psi & (n \geq 2) \\
&= i\bar{\psi} \left\{ -V\sum_{m=1}^{1-n} V^{-m} K_\mu(x) V^{n+m-2}(V-1) - K_\mu(x) V^{n-1} \right\} \psi & (n \leq 0)
\end{aligned} \qquad (30)$$

The conserved currents and the transformations with which they are associated are actually mismatched. As was already noted in the paper by Kikukawa and Yamada [7], the same current is associated with each of the two asymmetric forms of the Lüscher axial transformation, which we have called $\delta_A^{(0)}$ and $\delta_A^{(1)}$. Similarly, the same vector current is associated with each of the two transformations $\delta_V^{(0)}$ and $\delta_V^{(1)}$. The equalities of these pairs of currents is equivalent to our earlier observation that the differences $\left(\Delta_A^{(0)} - \Delta_A^{(1)}\right)$ and $\left(\Delta_V^{(0)} - \Delta_V^{(1)}\right)$ generate local gauge symmetries, because the Noether procedure gives zero when applied to a local symmetry.

*Ward Identities:* The Euclidean space analogues of conservation laws are the Ward identities. They express the physical implications of symmetries. They are derived from the path integral expressions for Green's functions (suppressing all space-time labels and the overall normalizations)

$$\langle \psi\psi\psi \cdots \bar{\psi}\bar{\psi}\bar{\psi} \cdots \rangle = \int \mathcal{D}\psi \mathcal{D}\bar{\psi} \, \psi\psi\psi \cdots \bar{\psi}\bar{\psi}\bar{\psi} \cdots e^{-S_F} \qquad (31)$$

by making a $\Delta_A^{(n)}$ or $\Delta_V^{(n)}$ transformation on the fermionic variables under the path integral. Inside the path integral, these transformations are just changes of integration variable, so do not affect its value. The first order transformations give

$$\int \mathcal{D}\psi \mathcal{D}\bar{\psi} \left(\Delta_{A,V}^{(n)} - Q_{A,V}^{(n)}\right)\left(\psi\psi\psi \cdots \bar{\psi}\bar{\psi}\bar{\psi} \cdots e^{-S_F}\right) = 0 \tag{32}$$

The $Q_{A,V}^{(n)}$ term comes from the Jacobean of the transformation generated by $\Delta_{A,V}^{(n)}$, which is the path integral formulation of the anomaly:

$$\begin{aligned} Q_A^{(n)}{}_{\text{singlet}} &= \sum_x \varepsilon(x) q_A^{(n)}(x) \\ Q_A^{(n)}{}_{\text{flavored}} &= 0 \\ Q_V^{(n)}{}_{\text{all components}} &= 0 \end{aligned} \tag{33}$$

Explicitly, the non-vanishing anomaly is

$$q_A^{(n)}(x) = Tr\, I(x)\gamma_5\left(V^n + V^{n-1}\right) \tag{34}$$

Using Eqs. (20) and (22) to express the effect of the transformation on each fermion variable and on $S_F$, and collecting the coefficients of $\varepsilon(x)$, give the Ward identities. For example, from the fermion propagator $\langle \psi \bar{\psi} \rangle$ we find the basic Ward identities

$$\begin{aligned} \gamma_5 I(x)V^n \langle \psi \bar{\psi} \rangle + \langle \psi \bar{\psi} \rangle V^{1-n} I(x)\gamma_5 + \partial_\mu^{(-)}\langle \psi J_\mu^{5(n)}(x) \bar{\psi} \rangle - q_A^{(n)}(x)\langle \psi \bar{\psi} \rangle &= 0 \\ -i I(x)V^n \langle \psi \bar{\psi} \rangle + i\langle \psi \bar{\psi} \rangle V^n I(x) \quad + \partial_\mu^{(-)}\langle \psi J_\mu^{(n)}(x) \bar{\psi} \rangle &= 0 \end{aligned} \tag{35}$$

The form of these identities, and especially their generalizations to higher Green's functions, show the connection with the conservation laws in canonically quantized field theory.

Not unexpectedly, all the flavor singlet axial symmetries are anomalous. However, except for $n = 0,1$, the values of their anomalies are different from the index of $D = 1 - V$, because they come from the Jacobeans of different transformations than $\Delta_A^{(1,0)}$.

Note that because of the cyclic symmetry of the trace, $q_A^{(n)}(x) = q_A^{(1-n)}(x)$.

*Illustrative Example of the Continuum Limit:* The problem of too many symmetries is generic to Ginsparg-Wilson fermions, and so we can instructively examine it in a solvable example — free overlap fermions. Even in this almost trivial case, the continuum limit turns out to be different, and subtler, than in other lattice theories. For this example we keep the lattice spacing explicit and allow for different lattice spacings along different directions. In momentum space, using the customary notation,

$$\hat{p}_\mu = \frac{1}{a_\mu} \sin(a_\mu p_\mu)$$
$$\hat{\hat{p}}_\mu = \frac{2}{a_\mu} \sin\left(\frac{a_\mu p_\mu}{2}\right)$$
$$M = \left(\sum_{\mu=1}^4 \frac{a_\mu}{2} \hat{\hat{p}}_\mu^2\right) - s \tag{36}$$
$$s \in \left(0, \frac{2}{\max_\mu a_\mu}\right)$$

the Dirac kernel is

$$aD = 1 - V = 1 + \frac{D_W}{\sqrt{D_W^\dagger D_W}} = \frac{i\gamma \cdot \hat{p} + M + \sqrt{\hat{p}^2 + M^2}}{\sqrt{\hat{p}^2 + M^2}} \tag{37}$$

where $a_\mu$ is the lattice spacing in the $\mu$ direction and the parameter $a$ with no subscript also has dimensions of length and will be chosen later to adjust the normalization of the fermion propagator. Eq. (37) of course satisfies the Ginsparg-Wilson equation and all the symmetry generators and Noether currents can be displayed explicitly. In this form we may take the continuum limit in each direction independently. If the spacing in direction $\mu$ goes to zero, then $\hat{p}_\mu \to p_\mu$, and the $\hat{\hat{p}}_\mu$ term drops out from $M$.

The key point to notice is that even along the continuous directions $\mu$, those for which $a_\mu = 0$, a $p_\mu^2$ term remains under the square root in the denominator. This implies that so long as the lattice spacing remains non-zero in any direction, the Dirac kernel remains non-local in all directions. In particular, if we take the limit of continuous time ($a_4 \to 0$) but keep the spatial lattice discrete, we do not obtain a Hamiltonian theory, because the resulting Dirac kernel is non-local in time. This non-locality is not just a failure of lattice ultralocality; it is real, physical non-locality. The scale of non-locality in the continuous time direction is set by the finite discrete spatial lattice spacing. The Dirac kernel continues to satisfy the Ginsparg-Wilson relation, and so all the extra symmetries and other problems discussed above are still present. Note that once we make time continuous we are free to continue to Minkowski space, but the non-locality and extra symmetries remain as in Euclidean space. This situation is a stark contrast with other Euclidean lattice actions, where taking the continuous-time limit directly gives the canonical Hamiltonian [8, 9].

In fact, even when all the spacings are taken to zero, so that the only remnant of the Wilson term is $M \to -s$, the Dirac kernel remains non-local

$$aD \to \frac{i\gamma \cdot p - s + \sqrt{p^2 + s^2}}{\sqrt{p^2 + s^2}} \tag{38}$$

and still invariant under all the unphysical extra Ginsparg-Wilson symmetries. In this example, the usual procedure for getting Minkowski-space Green's functions from a Euclidean-space lattice does not reach the target theory, or in fact any local theory.

The magnitude of the non-locality in (38) is controlled by $s$, the negative mass term whose value, so long as it is kept within the range indicated in Eq. (36), is regarded as having no physical significance; rather being just an "extra knob" with which to tune algorithms. In fact, since all lattice spacings have been taken to zero, $s$ can have any positive value, and that is the key to recovering ordinary free fermions. If we take the $s \to \infty$ limit and choose the normalization factor to be $a = 1/s$, then we arrive at the free massless Dirac kernel:

$$D \to i\gamma \cdot p \qquad (39)$$

It is only after this concatenation of limits that the extra symmetry structure we found in the first section of this note is suppressed, and even in this simplest of examples accomplishing it required reverting from the chiral symmetry of the Ginsparg-Wilson relation to ordinary continuum chiral symmetry.

*Concluding Remarks:* The foregoing observations have uncovered several signs that the connection between Ginsparg-Wilson lattice chiral fermions and continuum chiral fermions is problematic. Let us summarize them:

1) The chiral symmetry group of Ginsparg-Wilson fermions is not the same as the continuum chiral symmetry group. There are an infinite number of lattice chiral transformations for each continuum transformation, and just taking the continuum limit and analytically continuing to Minkowski space does not eliminate this unphysical superabundance of conserved quantities.

2) The connection between conserved charges and symmetry generators is (partially) lost. The generators of two pairs of different transformations, ($\delta_A^{(1)} \neq \delta_A^{(0)}$ and $\delta_V^{(1)} \neq \delta_V^{(0)}$) have the same Noether currents ($J_\mu^{5(0)} = J_\mu^{5(1)}$ and $J_\mu^{(0)} = J_\mu^{(1)}$). This is obviously incompatible with canonically quantized field theory, where the generator of a symmetry transformation is the space integral of the time component of its conserved current. Again, taking the continuum limit and going to Minkowski space does not automatically restore the necessary one-to-one correspondence between currents and symmetries.

3) The problems of extra chiral transformations and a mismatch between symmetries and currents can arise because the Euclidean space path integral is not canonical, and fermion and anti-fermion variables must be treated as independent, not as conjugate variables. However, this is certainly not the source of the problem: such pathologies are not a general feature of Euclidean fermionic path integrals.

In this note we have described a number of evidently related conflicts between the Ginsparg-Wilson treatment of lattice fermions and the properties of the target continuum theory. These conflicts are inherent in all implementations of lattice fermions based on the Ginsparg-Wilson relation, including overlap, domain-wall, and perfect-action chiral fermions. We have no solution to these severe problems with an otherwise very attractive formulation of lattice fermions.

**Acknowledgments**

The author would like to thank Steve Sharpe for discussions and advice about this work, and Wolfgang Bietenholz and Daniel Nogradi for their comments on the first version of this manuscript.

9 *Cf.* M. Creutz, I. Horváth, and H. Neuberger, Nucl. Phys. **B** (Proc. Suppl.) **106**, 760 (2002) [hep-lat/0110009]; K. Matsui, T. Okamoto, and T. Fujiwara, Phys. Rev. **D71**,114501 (2005.) [hep-lat/0501008]. The Hamiltonian theories discussed in these papers are not obtained as the continuous time limit of lattice actions.